\pdfoutput=1

\documentclass[11pt]{article}

\usepackage[preprint]{acl}

\usepackage{times}
\usepackage{latexsym}
\usepackage{multirow}
\usepackage[normalem]{ulem}
\useunder{\uline}{\ul}{}
\usepackage{algorithm}
\usepackage{algorithmic}
\usepackage{amsmath}
\usepackage{xcolor}
\usepackage{tcolorbox}
\usepackage{hyperref}

\definecolor{green}{RGB}{42, 126, 2}
\definecolor{light_red}{RGB}{234, 153, 153}
\definecolor{light_yellow}{RGB}{255, 242, 204}
\definecolor{light_orange}{RGB}{252, 229, 205}
\definecolor{light_blue}{RGB}{50, 150, 232}
\definecolor{light_gray}{gray}{0.3}
\definecolor{light_purple}{RGB}{203, 181, 255}
\definecolor{custom_purple}{RGB}{78, 0, 255}
\definecolor{custom_cyan}{RGB}{0, 120, 122}

\usepackage[T1]{fontenc}

\usepackage[utf8]{inputenc}

\usepackage{microtype}

\usepackage{inconsolata}

\usepackage{graphicx}
\usepackage{makecell}
\usepackage{enumitem}

\title{ReasoningRec: Bridging Personalized Recommendations and Human-Interpretable Explanations through LLM Reasoning}

\author{Millennium Bismay \\
  Texas A\&M University \\
  College Station, TX \\
  \texttt{mbismay@tamu.edu} \\
  \And
  Xiangjue Dong \\
  Texas A\&M University \\
  College Station, TX \\
  \texttt{xj.dong@tamu.edu} \\
  \And
  James Caverlee \\
  Texas A\&M University \\
  College Station, TX \\
  \texttt{caverlee@tamu.edu} \\}

\begin{document}
\maketitle
\begin{abstract}
This paper presents ReasoningRec, a reasoning-based recommendation framework that leverages Large Language Models (LLMs) to bridge the gap between recommendations and human-interpretable explanations. In contrast to conventional recommendation systems that rely on implicit user-item interactions, ReasoningRec employs LLMs to model users and items, focusing on preferences, aversions, and explanatory reasoning. The framework utilizes a larger LLM to generate synthetic explanations for user preferences, subsequently used to fine-tune a smaller LLM for enhanced recommendation accuracy and human-interpretable explanation. Our experimental study investigates the impact of reasoning and contextual information on personalized recommendations, revealing that the quality of contextual and personalized data significantly influences the LLM's capacity to generate plausible explanations. Empirical evaluations demonstrate that ReasoningRec surpasses state-of-the-art methods by up to 12.5\% in recommendation prediction while concurrently providing human-intelligible explanations. The code is available \href{https://anonymous.4open.science/r/reasoningrec-4A69}{here}.
\end{abstract}

\section{Introduction}
\label{section:introduction}


In our day-to-day life, many of our experiences are driven by word-of-mouth recommendations. For example, a friend may suggest a good book to read, knowing my preference for historical fiction and thriller novels. A work colleague may introduce me to a new software library based on our shared work history. These recommendations often focus on \textcolor{green}{likes} and \textcolor{red}{dislikes} of a user and provide \textcolor{custom_purple}{explanatory reasoning} for why a user may like a target item as in this example:

\begin{tcolorbox}[colback = white, colframe = custom_cyan!75!white]
\label{example:reasoningrec_output}
\small
   You \textcolor{green}{prefer movies which are heartwarming, inspiring, ... themes of love, family, and personal growth} ...  but \textcolor{red}{dislike movies that rely on predictable gags and tired humor} ... \textcolor{custom_purple}{You might like `My Dog Skip' as it is a heartwarming coming-of-age film about a young boy and his loyal dog, filled with friendship, adventure, and personal growth ...}
\end{tcolorbox}

In practice, many modern recommendation systems aim to scale recommendations based on the actions of millions of users, e.g., \cite{sasrec, caser, mf, ncf, lrurec, bert4rec}. These approaches, however, are typically not driven by such explicit human-interpretable \textcolor{custom_purple}{explanatory reasoning}, since such detailed reasons are rarely provided by users and challenging to infer from online system traces. Instead, these systems model implicit user-item interaction signals (e.g., from clicks or likes) toward predicting the next item a user may like. 

With the rise of Large Language Models (LLMs), we are motivated to revisit the core word-of-mouth reasoning assumption that inspired much of the early work in recommendation systems, e.g., \cite{social_info_filtering}, but that has been mostly missing from today's large scale recommender systems. That is, \textit{can we leverage the reasoning capabilities of LLMs to bridge the gap between recommendations and the human-intelligible, interpretable reasoning behind them?} Concretely, we propose a reasoning-based recommendation framework called \textit{ReasoningRec}. The core idea is to model users and items through LLM-powered generation (corresponding to the \textcolor{green}{likes} and \textcolor{red}{dislikes} above), and then use a large LLM to generate synthetic explanations for why a user may like an item (corresponding to the \textcolor{custom_purple}{explanatory reasoning}); these explanations are then used to fine-tune a smaller LLM (SLM) toward generating more accurate recommendations than methods that ignore such powerful reasoning chains. A side effect of the model is that it can also generate such human-interpretable explanations alongside its predictions. 

In our experiments, we find that the ability of the LLM to generate plausible explanations is heavily influenced by the quality of the contextual and personalized information about both users and items. That is, simply relying on an LLM to reason about items in the absence of such rich contextual information results in poor performance. Instead, finetuning according to the proposed framework, ReasoningRec, with such rich contextual and personalized information about users and items, results in improved performance in recommendation prediction while generating human-interpretable and intelligible explanations. We further observe that user profiles play a significant role for long user-item interaction sequences while item descriptions play a crucial role for datasets with high sparsity.

In summary, our contributions are as follows:

\begin{itemize}
[noitemsep,topsep=5pt,itemsep=2pt, leftmargin=20pt]
    \item We present the first comprehensive study on  the impact of human-interpretable explanations through LLM generated reasoning using contextual information, in the context of personalized recommendation.
    \item We propose ReasoningRec, a robust, efficient and effective instruction fine-tuning framework to enhance the performance of SLMs in both recommendation prediction and reasoning generation.
    \item We conduct extensive experiments and show that ReasoningRec outperforms the state-of-the-art (SOTA) up to $12.5\%$ on recommendation prediction, while also introducing a novel dimension to recommendation tasks by bridging the gap between recommendations and human-intelligible reasoning.
\end{itemize}

\begin{figure*}[]
  \includegraphics[width=\linewidth]{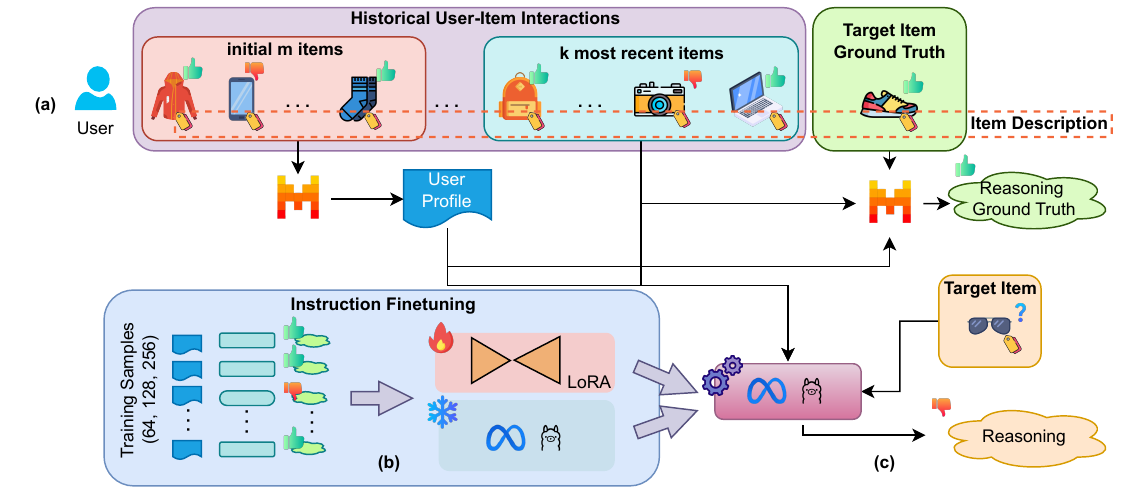} \hfill
  \caption {\textbf{Overview of ReasoningRec framework.} (a) We use an LLM, \textit{Mixtral-8x7B-Instruct-v0.1}, to generate \textit{item description} and \textit{user profile}. We use Chain-of-Thought prompting with the most recent $k$ items and leverage the rich semantic information to generate \textit{reasoning ground truth}, conditioned on the target item ground truth rating. (b) We propose a lightweight instruction finetuning framework, utilizing LoRA architecture, with very limited training samples to instruction finetune an SLM, \textit{Llama-2-7b-chat-hf}. (c) We use the instruction finetuned model to perform recommendation prediction and reasoning generation.}
  \label{fig:ReasoningRec}
\end{figure*}

\section{Related Work}
\label{section:related_works}

\paragraph{\textbf{LLMs for Recommender Systems.}} 
Recent advancements in LLMs have introduced transformative capabilities across a broad range of applications, including recommender systems. Influenced by the In-Context Learning (ICL) abilities \cite{icl_survey} of LLMs and their capacity to follow Chain of Thought (CoT) prompting \cite{cot}, initial work explored task-specific prompting paradigms to tailor LLMs for downstream tasks such as re-ranking~\cite{exranker} and conversational recommender systems~\cite{unicrs}. Some methods have further investigated the zero-shot capability of LLMs to understand user preferences \cite{kang2023llms}. More recently, research has shifted towards cost and compute-effective instruction fine-tuning methods of Small Language Models (SLMs) for predicting binary recommendation ratings, as seen in TALLRec \cite{tallrec}, PALR \cite{palr}, and ReLLa/ReiT \cite{rella}. This fine-tuning process involves training pre-trained models using task-specific recommendation datasets, which include user-item interactions (e.g., purchases, ratings, clicks). These finetuned SLMs demonstrate improved performance compared to LLMs relying solely on zero-shot prompting.

\paragraph{\textbf{Explanations in Recommendation.}} Prior work has aimed to explain or justify what items are suggested by a recommender system, e.g., \cite{justifying, learning-generate, li-etal, ni-etal}. These methods typically generate explanations post-hoc, and are not incorporated for training new recommendation models. In this work, we propose to generate synthetic explanations based on user-item interactions that can then be used to fine-tune a more effective recommender.

\paragraph{\textbf{LLM Reasoning.}} 
Recent studies have suggested that LLMs with parameters beyond a certain scale exhibit emergent reasoning abilities \cite{emergent, mwp1}. When provided with examples of ``chain of thought'' (CoT), which represent intermediate natural language reasoning steps, these models can generate explicit rationales similar to eliciting inductive/abductive reasoning \cite{cot, reasoning_survey}. Advances such as zero-shot CoT \cite{zeroshot_cot}, where a model is prompted with phrases like ``Let's think step by step,'' allow for reasoning without the need for explicit few-shot examples. Building on this, multi-step reasoning approaches -- such as Successive Prompting \cite{successive}, Tree-of-Thought \cite{tot}, Graph-of-Thought \cite{got, got2}, Iterating CoT \cite{iterative_cot}, and Self-Consistency \cite{self_consistency} -- have become critical strategies for enhancing reasoning in various downstream tasks. In the recommendation domain, Logic-Scaffolding \cite{logic_scaffolding} provides a preliminary CoT framework for generating reasoning. Additionally, there has been an early exploration into knowledge adaptation using multiple experts in cross-domain knowledge transfer \cite{kar}. Despite these advancements, reasoning-based approaches for recommendation prediction remain underexplored. A notable exception is Rec-SAVER \cite{recsaver}, a recent approach which applies reasoning generation for rating prediction by fine-tuning an SLM. 

\section{ReasoningRec}
\label{section:reasoningrec}
In this section, we introduce the framework of ReasoningRec. We begin with preliminaries, then introduce how we model items and user profiles, before creating reasoning-based explanations that guide our model fine-tuning.

\subsection{Preliminaries}
\label{section:preliminaries}
We define the recommendation task as a binary classification problem using multi-field categorical data. Given a dataset $\mathcal{D} = \{(X_n, y_n)_{n=1}^N\}$, where $X_n$ represents the user features (e.g., User ID, gender, age) and item features (e.g., Item ID, title, brand) for the $n$-th instance, and $y_n$ denotes the corresponding binary label (with $y_n=1$ indicating a ``like'' and $y_n=0$ indicating a ``dislike''), the goal of the recommendation system is to learn a function $\mathcal{F}(\cdot)$ that accurately predicts the probability of a ``like'' for each sample $X_n$, i.e., $P(y_n = 1|X_n)$. The predicted probability $\hat{y}_n$ is given by $\hat{y}_n = \mathcal{F}(X_n;\theta)$, where $\theta$ represents the model parameters.

Let $\mathcal{U} = \{u_1, u_2, \dots, u_{|\mathcal{U}|}\}$ denote the set of users and $\mathcal{I} = \{i_1, i_2, \dots, i_{|\mathcal{I}|}\}$ denote the set of items. The historical interaction sequence of a user $u\in\mathcal{U}$ is represented as $\mathcal{S}^u = (s_1^u, s_2^u, \dots, s_{|\mathcal{S}^u|}^u)$, where $s_i^u$ refers to the chronological interaction of user $u$ with item $i$, $\forall i \in \mathcal{I}$. Each user-item interaction is represented by a rating, $r'_{u,i} \in [1, 5]$, and an optional review $K_{u,i}^*$\footnote{$K_{u,i}^*$ is unavailable for the ML1M dataset and sparsely available for the Beauty and Fashion datasets}, which is a textual string representing the user's feedback for the item. To adapt the ratings into a binary classification framework, we binarize the rating using a threshold $r_\mathcal{T}$. Specifically, we  define $r_{u,i} = 1$ if $r'_{u,i} > r_{\mathcal{T}}$ and $r_{u,i} = 0$ otherwise.


\subsection{Item Description Generation}
\label{section:item_description_generation}
We aim to model user preferences based on past interactions with items. Hence, it is crucial to create human-intelligible item descriptions with rich semantic information to enhance these models of user preference.
Formally, we define the item description $D_i$ for an item $i$ as:
 \begin{equation*}
    \label{eq:item_description}
    D_i = \mathcal{L}_M(\mathcal{M}_i),
\end{equation*}
where $i \in \mathcal{I}$, and $\mathcal{M}_i$ is the metadata corresponding to that item, consisting of details such as -- title, genre, brand, price, and reviews. For each item $i$, we curate a list of at most $p$ user reviews. We use an LLM, $\mathcal{L}_M$, to generate a concise $n$-word item description -- $D_i$, which takes into account the item metadata $\mathcal{M}_i$ and the selected item reviews. Appendix \ref{section:appendix_item_desc} provides the item description generation process in details. Appendix~\ref{section:appendix_prompts} Table~\ref{tab:item_description_prompt} shows the prompts used to create the item descriptions.
\begin{tcolorbox}[colback = white, colframe = custom_cyan!75!white]
\label{example:item_description}
\small
\textcolor{blue}{The product is a beautiful watch, with big size ...}. \textcolor{green}{It is comfortable, easy to put on/take off, perfect for workouts} but \textcolor{red}{has a short-lived rubber band.}
\end{tcolorbox}
As shown in the above example, for an item, the blue text provides a concise description, the green text describes the positive features of the item, and the red text describes the negative features.

\subsection{User Profile Creation}
\label{section:user_profile_creation}
Sequential recommendation tasks typically consider the most recent $k$ items to predict whether the user will like or dislike the next item. The idea behind this approach is to capture temporal user behavior from the most recent interactions. However, this process overlooks some of the user's broader preferences which might have been exhibited prior to the most recent $k$ items.  To address this, 
we create a user profile $P_u$, a concise $q$-word profile summary that captures the initial behavior and preference of a user $u\in\mathcal{U}$. To create this user profile, we select at most the first $m$ user interacted items such that the set of items does not overlap with the most recent $k$ items. This ensures that the user profile reflects diverse preferences from earlier interactions. For each of the $m$ items, we use $r_{u,i}$ to assess whether the user liked the item or not. We then prompt an LLM, $\mathcal{L}_M$, with the $r_{u,i}$ and corresponding item description, $D_i$, following the Chain-of-Thought (CoT) \cite{cot} prompts as shown in Appendix~\ref{section:appendix_prompts} Table~\ref{tab:user_profile_prompt} to generate a 100-word user profile, $P_u$, summarizing the user preferences. Formally,

\begin{equation*}
    \label{eq:user_profile}
    P_u = \mathcal{L}_M((s_i^u)_{i=1}^{m}),
\end{equation*}

\noindent where $s_i^u= <r_{u,i}, D_i>$,  represents the user-item interactions for the user $u$ with item $i$.

\begin{tcolorbox}[colback = white, colframe = custom_cyan!75!white]
\label{example:user_profile}
\small
\textcolor{green}{The user seems to prefer fashionable, eye-catching products at a reasonable price...} However, \textcolor{red}{they dislike products that have quality issues such as loose fit ...}
\end{tcolorbox}
As shown in the above example, the green text represents the features \textcolor{green}{liked} by the user, and the red text represents the features \textcolor{red}{disliked} by the user.

\begin{table*}
  \centering
  \resizebox{\linewidth}{!}{
  \begin{tabular}{p{2cm}p{15cm}}
    \hline
    \hline
    &   \textbf{Prompt Template and Output for ReasoningRec} \\
    \hline
    &   \textcolor{orange}{You are an expert \textit{<assign\_role\_for\_llm>}.}You are provided with the user profile and list of recent \textit{<items>} that the user has \textit{<user\_activity>} and whether the user likes it or not.\\
    &   User Profile -- \textcolor{blue}{<$P_u$>}\\
    &   List of recent \textit{<items>} and their description -- \textcolor{light_blue}{\textit{Liked/Disliked} $<i_, D_i>_{i = t-k-1}^{t-1}$}\\
    &   \textcolor{gray}{Analyze all the information given in order. Do not use any information not mentioned above.}\\
    Prompt  &   \textcolor{purple}{Predict whether the user will like the target \textit{<item>} -- <$i_t, D_{i_t}$> or not. Answer with a Yes or No in the following format -- Prediction: Yes or No by analyzing the user's behavior from the given list of \textit{<items>} and identify the \textit{<item>} characteristics that the user likes and dislikes in at most 100 words. Explain with reasoning whether the user will like or dislike the target \textit{<item>} -- <$i_t$> in at most 100 words.} \\
    \\
    &   \textcolor{custom_cyan}{Prediction: Yes}\\
    &   \textcolor{green}{The user has shown a preference for cute and comfortable accessories ...}\\
    Output  &   \textcolor{red}{They have disliked products that appear cheap or break easily ...}\\
    &   \textcolor{custom_purple}{... the bracelet is described as high quality and nicely polished, suggesting it will not have these issues ... the user is likely to appreciate the bracelet's adjustable and high-quality features ...}\\

    \hline
  \end{tabular}}
\caption{\label{tab:reasoning_rec_prompt_template}
   \textbf{\textit{Prompt given to LLM}} ($\mathcal{L}$) -- \textcolor{orange}{We first assign a role to the llm}. We provide the LLM with \textcolor{blue}{User Profile, $P_u$,} and \textcolor{light_blue}{Item Descriptions, $D_i$, of k most recent interacted items and their corresponding preference, $r_{u,i}$.} \textcolor{gray}{We assign some guardrails to the LLM to mitigate hallucinations}. Finally, for a target item $D_{i_t}$, we provide a  \textcolor{purple}{Chain-of-Thought instruction to predict and generate reasoning.} \textbf{\textit{Output}} -- \textcolor{custom_cyan}{The model first generates the recommendation prediction, $\hat{r}_{u,i_t}$}. It mentions some key features that are \textcolor{green}{liked} or \textcolor{red}{disliked} by the user. Finally, \textcolor{custom_purple}{the model generates the reasoning for the provided prediction, $\hat{R}_{u,i_t}$}.
  }
\end{table*}

\subsection{Explanatory Reasoning Generation}
\label{section:reasoning_groundtruth_generation}
In this section, our goal is to create explicit human-interpretable \textcolor{custom_purple}{explanatory reasoning} to guide our recommendation framework. Recall that most user-item interactions are implicit and so explicit reasoning about why a user prefers a particular item is typically unavailable. Even for the rare cases where a user may explicitly indicate a preference (as in the case of a written review), patterns connecting sequential interactions are often missing. 



We employ a well-curated zero-shot CoT prompting method to generate synthetic \textcolor{custom_purple}{explanatory reasoning} for a user's implicit preference towards the target item, by conditioning the generation process on the user's implicit feedback for the target item. This process is crucial to the quality of human-intelligible reasoning generation. For a user $u$ and target item $i_t$, we utilize the User Profile $P_u$, recent $k$ user-item interactions for the same user $u$, $(s_i^u)_{i=t-k-1}^{t-1} = <r_{u,i}, D_i>_{i = t-k-1}^{t-1}$, and ground truth rating $r_{u,i_t}$ which indicates whether the user originally \textcolor{green}{liked} or \textcolor{red}{disliked} the target item. This information is used to generate reasoning $R_{u,i_t}$ which explains the user's behavior about what features are most likely to influence the user's decision, given the ground truth rating $r_{u,i_t}$. We formally define this process as:
\begin{equation*}
  \label{eq:response_generation}
  R_{u,i_t}  = \mathcal{L}_M(P_u, (s_i^u)_{i=t-k-1}^{t-1} | r_{u,i_t}).
\end{equation*}
The structure of the prompts is detailed in Appendix~\ref{section:appendix_prompts} Table~\ref{tab:response_generation_prompts}.

Note that such \textcolor{custom_purple}{explanatory reasoning} is based on previous user-item interactions (where we know the ground truth of whether the user likes an item or not). It is not typically available in practice since the model must predict whether a user will like an item or not. Hence, we next demonstrate how to fine-tune a smaller LLM  over these synthetic reasoning chains toward generating more accurate recommendations.

\subsection{Finetuning with Personalized Reasoning}
\label{section:instruction_finetuning}

In this section, we demonstrate our framework, ReasoningRec, for instruction finetuning the SLM, \textit{Llama-2-7b-chat-hf}, denoted as $\mathcal{L}_L$. Let $\Theta$ be the pre-trained parameters of $\mathcal{L}_L$. Our approach is based on the conditional language model objective. However, finetuning an LLM can be both computationally intensive and highly time-consuming, particularly for long sequences. To address this, we adopt LoRA (Low-Rank Adaptation)~\cite{lora}, which involves freezing the pre-trained model's parameters $\Theta$ and introducing trainable low-rank decomposition matrices into select layers of the Transformer architecture. It is possible to achieve near-equivalent performance to full-model finetuning by updating only a small subset of LoRA parameters~\cite{lora}, designated as $\Phi$. By optimizing these low-rank matrices, we can efficiently and effectively incorporate new information without altering the frozen parameters. The final learning objective is computed as:
\begin{equation*}
  \label{eq:finetuning_objective}
\max_{\Phi} \sum_{(p,q) \in \mathcal{D}_T} \sum_{v=1}^{|q|} \log \left( \mathcal{P}_{\Theta + \Phi} (q_v \mid p, q_{<v}) \right),
\end{equation*}
where $\Theta$ represents the pre-trained frozen parameters of the LLM $\mathcal{L}_L$, and $\Phi$ is the LoRA parameters trained and updated during finetuning. $\mathcal{D}_T$ is the Training dataset, consisting of $\mathcal{K}$ pairs of $(p,q)$, where $p$ is the ``Instruction Input'' and $q$ is the ``Instruction Output''. Let $q_v$ be the $v$-th token of the output $q$, and $q_{<v}$ represents all the tokens before $q_v$. We define $p$ as an instruction prompt comprising the User Profile $P_u$, recent $k$ user-item interactions for the same user $u$, $(s_i^u)_{i=t-k-1}^{t-1} = <r_{u,i}, D_i>_{i = t-k-1}^{t-1}$, Item description for target item $D_{i_t}$ and an instruction to generate $q$, where $q$ consists of the binary ground truth rating $r_{u,i_t}$ and a ground truth reasoning $R_{u,i_t}$. As we generated $R_{u,i_t}$ in Section~\ref{section:reasoning_groundtruth_generation}, based on the ground truth rating $r_{u, i_t}$, we use this as the ground truth reasoning for training the model. The goal is to train $\mathcal{L}_L^{\Theta + \Phi}$ in such a way that, 
\begin{equation*}
  \label{eq:finetune_generation_objective}
  <\hat{r}_{u,i_t}, \hat{R}_{u,i_t}>  = \mathcal{L}_L^{\Theta + \Phi}(P_u, (s_i^u)_{i=t-k-1}^{t-1}, D_{i_t}),
\end{equation*}
where $\hat{r}_{u,i_t}$ represents the binary rating prediction of user $u$ for target item $i_t$ and $\hat{R}_{u,i_t}$ represents the generated reasoning describing the user behavior and factors affecting the user's predicted rating. The Instruction Input Template and Output example are detailed in Table~\ref{tab:reasoning_rec_prompt_template}. The actual prompts are detailed in Appendix~\ref{section:appendix_prompts} Table~\ref{tab:reasoning_rec_prompt}.

\section{Experimental Setup}
\label{section:experiments_and_results}
In this section, we introduce the datasets, baseline methods, evaluation metrics, and implementation details used in our experiments.



\subsection{Datasets}
\label{section:experimental_setup}
Given that real-world datasets frequently exhibit different levels of sparsity, we aim to evaluate the performance of our methods across datasets with varying degrees of sparsity. 

\textit{\textbf{ML1M}}~\cite{ml1m} is a widely used benchmark dataset about movies with 1 million user-item interactions.

\smallskip
\textit{\textbf{Amazon Fashion}} and \textit{\textbf{Beauty}}~\cite{justifying} are series of datasets comprising large corpora of product reviews crawled from Amazon. They have high sparsity.

Following the SASRec~\cite{sasrec} approach, for a user-item historical sequence, we use the last interaction as test data, the second last interaction as validation data, and the third last interaction as training data. Details about datasets are shown in Appendix~\ref{section:appendix_dataset}.

\subsection{Baselines and Evaluation Metrics}
\label{section:baselines}
We consider two representative sequential recommenders -- \textbf{\textit{SASRec}}~\cite{sasrec} and \textit{\textbf{Caser}}~\cite{caser}. These are pre-LLM recommenders that are trained over historical user-item interactions. \textbf{\textit{SASRec}} is a strong attention-based sequential recommender and \textbf{\textit{Caser}} is a CNN-based sequential recommender.


\textit{\textbf{TALLRec}}~\cite{tallrec} is an LLM-based recommender that uses an instruction finetuning method with $\mathcal{K}\in\{64, 128, 256\}$ training samples to finetune an SLM, \textit{Llama-2-7b-chat-hf}  for recommendation prediction.

\textit{\textbf{Rec-SAVER}}~\cite{recsaver} is a contemporaneous approach that addresses the simultaneous tasks of rating prediction and reasoning generation by finetuning a \textit{Flan-T5 XL} model. Specific details are mentioned in Appendix~\ref{section:appendix_dataset}.

 We also consider two zero-shot methods for comparison -- \textit{\textbf{Zero-shot Vanilla}} and \textbf{\textit{Zero-shot ReasoningRec}}. \textit{\textbf{Zero-shot Vanilla}} is prompted with only item titles for the recommendation prediction task. \textit{\textbf{Zero-shot ReasoningRec}} is prompted with all contextual and semantic information, including item descriptions and user profiles, to generate both recommendation predictions and corresponding reasoning. We follow the same prompt template shown in Table~\ref{tab:reasoning_rec_prompt_template} and use \textit{Llama-2-7b-chat-hf} and \textit{Mixtral 8x7b Instruct v0.1} as two backbone models for these two baselines.

Our study focuses on both recommendation prediction and the quality of reasoning generation. We employ \textit{\textbf{Binary AUC}}~\cite{auc} for evaluating the binary recommendation prediction task. We use \textit{\textbf{BERTScore}}~\cite{bertscore}, which uses contextual embeddings to capture the semantic similarity between the generated reasoning and the ground truth reasoning to access the quality of reasoning generation. Specific details are mentioned in Appendix~\ref{section:appendix_evaluation_metrics}

\subsection{Implementation Details}
\label{section:implentation_details}
 We binarize the ratings using the threshold $r_\mathcal{T} = 3$ for our experiments, following TALLRec~\cite{tallrec} paper. As mentioned in Section~\ref{section:item_description_generation}, for item descriptions, we use at most $p = 10$ reviews sampled in a stratified manner and generate a $n = 25$ word description for every item. For user profile, we use at most $m = 15$ initial items to generate a concise $q = 100$ word profile summary, capturing the user preference and behavior from their initial interactions, as mentioned in section~\ref{section:user_profile_creation}. For all the generation tasks, we use \textit{Mixtral-8x7b-Instruct-v0.1} as the LLM.

For the proposed method, \textbf{\textit{ReasoningRec}}, we instruction finetune an SLM, \textit{Llama2-7b-chat-hf}, with limited number of training samples, $\mathcal{K} \in \{64, 128, 256\}$, following TALLRec. We conduct all experiments on NVIDIA RTX A5000 24GB GPUs.
Additional implementation details for our methods, models, and finetuning are explained in Appendix~\ref{section:appendix_implementation_details}


\section{Results and Analysis}
\label{section:results_analysis}

\begin{table*}[h]
\centering
\resizebox{\linewidth}{!}{%
\begin{tabular}{cccccccccc}
\hline
\hline
\multirow{3}{*}{\textbf{Model}} & \multicolumn{3}{c}{\textbf{ML1M}} & \multicolumn{3}{c}{\textbf{Fashion}} & \multicolumn{3}{c}{\textbf{Beauty}} \\ \cline{2-10} 
 & $\mathcal{K}$ & \textbf{AUC} & \textbf{\makecell{BERT\\ Score}} & $\mathcal{K}$ & \textbf{AUC} & \textbf{\makecell{BERT\\ Score}} & $\mathcal{K}$ & \textbf{AUC} & \textbf{\makecell{BERT\\ Score}} \\ \hline
SASRec & 6040 & 0.594 & - & 1273 & 0.526 & - & 624 & 0.538 & - \\
Caser & 6040 & 0.619 & - & 1273 & 0.549 & - & 624 & 0.551 & - \\ \hline
Zero-shot Vanilla  & 0 & 0.509 & - & 0 & 0.473 & - & 0 & 0.475 & - \\
Zero-shot ReasoningRec  & 0 & 0.510 & - & 0 & 0.500 & - & 0 & 0.510 & - \\
Zero-shot Vanilla (\textit{Mixtral}) & 0 & 0.523 & - & 0 & 0.620 & - & 0 & 0.646 & - \\
Zero-shot ReasoningRec (\textit{Mixtral}) & 0 & 0.614 & 0.658 & 0 & 0.731 & 0.667 & 0 & 0.693 & 0.664 \\ \hline
\multirow{3}{*}{TALLRec } & 64 & 0.546 & - & 64 & 0.653 & - & 64 & 0.684 & - \\
 & 128 & 0.621 & - & 128 & 0.705 & - & 128 & 0.721 & - \\
 & 256 & 0.659 & - & 256 & 0.753 & - & 256 & 0.742 & - \\ \hline
Rec-SAVER XL& - & - & - & - & - & - & 4000 & {\ul 0.78} & 0.67 \\ \hline
\multirow{3}{*}{ReasoningRec  \textbf{(Ours)}} & 64 & 0.604 & 0.709 & 64 & 0.675 & 0.677 & 64 & 0.741 & 0.686 \\
 & 128 & {\ul 0.691} & {\ul 0.719} & 128 & {\ul 0.762} & {\ul 0.688} & 128 & 0.768 & {\ul 0.687} \\
 & 256 & \textbf{0.744} & \textbf{0.724} & 256 & \textbf{0.786} & \textbf{0.696} & 256 & \textbf{0.798} & \textbf{0.691} \\ \hline
\end{tabular}%
}
\caption{\label{tab:finetuned_reasoningrec}
Recommendation performance of ReasoningRec and baselines across three datasets. SASRec and Caser are trained on the complete dataset while TALLRec and our method use instruction finetuned SLM with $\mathcal{K} \in \{64, 128, 256\}$ samples. Zero-shot methods are used with both LLMs and SLMs to understand the importance of finetuning on SLMs. The best results are highlighted in \textbf{bold}, and the second best results are \uline{underlined}.}
\vspace{-10pt}
\end{table*}


\subsection{ReasoningRec outperforms all baselines}
\label{section:reasoningrec_vs_sota}

Table~\ref{tab:finetuned_reasoningrec} compares the performance of the proposed method, ReasoningRec, with baselines like SASRec~\cite{sasrec}, Caser~\cite{caser}, and TALLRec~\cite{tallrec}. We report the results from the contemporaneous Rec-SAVER XL paper~\cite{recsaver} for the Beauty dataset, based on a finetuned Flan-T5 XL model.


We see the highest performance gain on the ML1M dataset $(12.59\%)$, as compared to $4.38\%$ on Fashion and $7.55\%$ on Beauty, over TALLRec performance. The proposed ReasoningRec framework outperforms all other tested methods including traditional methods like SASRec, Caser, and finetuned SLM methods like TALLRec in recommendation prediction capability while providing human intelligible and interpretable reasoning validating the prediction. The finetuned ReasoningRec outperforms a contemporary work, Rec-SAVER XL, which leverages a highly compute-intensive training process by generating multiple reasoning outputs and selecting the best one for finetuning a Flan-T5 XL model, through self-verification using another prompt to the same LLM. It is also finetuned with a considerably higher number of training examples as compared to ReasoningRec. \textbf{\textit{Improved performance of ReasoningRec could be attributed to the use of rich semantic information such as User Profile and enriched Item Descriptions.}} The model is finetuned with only $\mathcal{K} = 256$ training samples and is able to outperform all other models in both recommendation prediction and reasoning generation, showcasing the cost and compute - efficiency, and effectiveness of the proposed lightweight instruction finetuning framework with LoRA~\cite{lora} architecture.


\subsection{ReasoningRec compared to Zero-shot larger LLMs}
\label{section:reasoningrec_vs_zeroshot}

We investigate the recommendation prediction and reasoning ability of an LLM (\textit{Mixtral 8x7b Instruct v0.1}), in a zero-shot setting -- not finetuned on any training sample, i.e. $\mathcal{K} = 0$. LLMs are pre-trained on a huge corpus of data with a very large number of parameters, which help them to perform well across various tasks in zero-shot settings using ICL abilities~\cite{icl_survey}. We use the CoT prompt\footnote{An ablation of Prompts is provided in Appendix~\ref{section:appendix_template_ablation}} as depicted in Table~\ref{tab:reasoning_rec_prompt_template}, which is used for the ReasoningRec framework and evaluated the results as shown in Table~\ref{tab:finetuned_reasoningrec}. \textbf{\textit{We demonstrate strong results for zero-shot ReasoningRec, outperforming traditional SoTA methods like SASRec and Caser}} and performing comparably to finetuned models with $\mathcal{K} = 64$ training samples. 

We also investigate the zero-shot capabilities of SLM (\textit{Llama2-7b-hf-chat}). We observe that SLMs perform poorly with recommendation tasks in a zero-shot setting. Zero-shot ReasoningRec improved the average prediction performance by $14.3\%$ from the Vanilla method\footnote{Vanilla: prompted with only item titles to predict the binary recommendation task}, however, the average performance of the SLM improved by only $3\%$, which underscores the drawback of SLMs in performing recommendation prediction even when provided with rich contextual information. However, once the same SLM is instruction finetuned according to the proposed framework, ReasoningRec, it outperforms the zero-shot ReasoningRec and every other baseline method. \textbf{\textit{Hence, we establish that the proposed lightweight instruction finetuning framework with synthetic reasoning ground truth helps in effectively improving the recommendation prediction as well as reasoning generation performance in SLMs.}}



\subsection{Ablation study for ReasoningRec}
\label{section:ablation_reasoningrec}

\begin{table*}[h]
\centering \small
\resizebox{\linewidth}{!}{%
\begin{tabular}{cccccccc}
\hline
\hline
\multirow{2}{*}{\textbf{Model}} & \multirow{2}{*}{$\mathcal{K}$} & \multicolumn{2}{c}{\textbf{ML1M}} & \multicolumn{2}{c}{\textbf{Fashion}} & \multicolumn{2}{c}{\textbf{Beauty}} \\ \cline{3-8} 
 &  & \textbf{AUC} & \textbf{BERTScore} & \textbf{AUC} & \textbf{BERTScore} & \textbf{AUC} & \textbf{BERTScore} \\ \hline
\multirow{3}{*}{ReasoningRec w/o Description} & 64 & 0.563 & 0.686 & 0.663 & 0.649 & 0.701 & 0.619 \\
 & 128 & 0.624 & 0.696 & 0.700 & 0.664 & 0.748 & 0.621 \\
 & 256 & 0.681 & 0.700 & 0.750 & 0.665 & 0.761 & 0.662 \\ \hline
\multirow{3}{*}{ReasoningRec w/o Profile} & 64 & 0.556 & 0.671 & 0.661 & 0.648 & 0.645 & 0.628 \\
 & 128 & 0.621 & 0.684 & 0.720 & 0.664 & 0.741 & 0.682 \\
 & 256 & 0.662 & 0.699 & {\ul 0.779} & 0.667 & {\ul 0.792} & 0.685 \\ \hline
\multirow{3}{*}{ReasoningRec} & 64 & 0.604 & 0.709 & 0.675 & 0.677 & 0.741 & 0.686 \\
 & 128 & {\ul 0.691} & {\ul 0.719} & 0.762 & {\ul 0.688} & 0.768 & {\ul 0.687} \\
 & 256 & \textbf{0.744} & \textbf{0.724} & \textbf{0.786} & \textbf{0.696} & \textbf{0.798} & \textbf{0.691} \\ \hline
\end{tabular}%
}
\caption{\label{tab:ablation_finetuned_reasoningrec}
\textbf{Ablation on ReasoningRec} - Understanding the importance of Item Description and User Profile. ReasoningRec w/o Description is finetuned with only user profiles to generate recommendation prediction and reasoning. Similarly, ReasoningRec w/o Profile is finetuned with only item descriptions.
}
\vspace{-10pt}
\end{table*}

Table~\ref{tab:ablation_finetuned_reasoningrec} demonstrates the importance of features like Item Description and User Profile on the recommendation prediction and reasoning generation capability. We demonstrate that item description plays a significant role in improving the quality of reasoning generation as ReasoningRec w/o Profile consistently outperforms ReasoningRec w/o Description. Specifically, in extremely sparse datasets like Beauty, item description plays a crucial role due to additional contextual information improving both reasoning and recommendation predictions. We establish that user profile is crucial for long interaction sequences such as in ML1M which signifies the importance of capturing the user's early preferences. We also show the collective importance of both profile and description for ReasoningRec which consistently outperforms the other methods. \textbf{\textit{We observe a positive correlation between the recommendation prediction performance (Binary AUC) with the quality of reasoning (BERTScore), which underscores the importance of reasoning for improvement in recommendation prediction performance.}}

\subsection{Ablation study for Zeroshot ReasoningRec}
\label{section:zeroshot_ablation}
We investigated the impact of key factors like Item Description and User Profile on the recommendation prediction and reasoning generation performance for zero-shot ReasoningRec. Table~\ref{tab:zero_shot_results} presents the ablation study of zero-shot methods.
\begin{table}[h]
\centering
\resizebox{\columnwidth}{!}{%
\begin{tabular}{cccc}
\hline
\hline
\multirow{2}{*}{\textbf{Method}} & \multirow{2}{*}{\textbf{\begin{tabular}[c]{@{}c@{}}ML1M\\ (AUC)\end{tabular}}} & \multirow{2}{*}{\textbf{\begin{tabular}[c]{@{}c@{}}Fashion\\ (AUC)\end{tabular}}} & \multirow{2}{*}{\textbf{\begin{tabular}[c]{@{}c@{}}Beauty\\ (AUC)\end{tabular}}} \\
 &  &  &  \\ \hline
Vanilla & 0.524 & 0.620 & 0.646 \\
w/ Description & 0.561 & 0.612 & 0.610 \\
w/ Profile & 0.585 & 0.641 & 0.618 \\
w/ Description, Profile & {\ul 0.591} & {\ul 0.646} & 0.651 \\
\hline
Reasoning & 0.562 & 0.593 & 0.594 \\
Reasoning w/ Description & 0.585 & 0.647 & 0.661 \\
Reasoning w/ Profile & 0.571 & 0.665 & {\ul 0.661} \\
\begin{tabular}[c]{@{}c@{}}Zeroshot ReasoningRec \\ Reasoning w/ Description, Profile\end{tabular} & \textbf{0.614} & \textbf{0.731} & \textbf{0.693} \\ \hline
\end{tabular}%
}
\caption{\label{tab:zero_shot_results}
   \textbf{ Ablation on Zero-shot Results:} Reasoning is effective with key factors such as User Profiles and Item Descriptions to enhance the recommendation prediction.
  }
  \vspace{-5pt}
\end{table}
 An initial analysis demonstrates that reasoning on item titles only performs worse than the Vanilla (No Reasoning, Only Binary Prediction) method, suggesting the model's inability to generate meaningful reasoning from limited inputs like item titles, reinforcing the importance of combining reasoning with rich semantic context. We see a gradual improvement in performance when we introduce Item Description and then User Profile into the input prompt. We assess that, LLMs inherently perform better at recommendation tasks with crucial semantic and contextual information such as Description and Profile. The zero-shot ReasoningRec method, which incorporates both Item Descriptions and User Profiles, consistently outperforms other zero-shot methods. \textbf{\textit{These results validate that LLM Reasoning is bolstered by the addition of rich semantic information such as User Profiles and Item Descriptions which personalizes the recommendations, making them more accurate, and contextually relevant, and boosts the recommendation prediction task.}}

\section{Conclusion}
\label{section:conclusion}
In this work, we studied the importance of rich contextual and semantic information about users and items in generating human-interpretable explanations. We introduced ReasoningRec, a lightweight and efficient framework, to instruction finetune an SLM on synthetic reasoning data generated by an LLM, conditioned on ground truth user ratings. We observed that our proposed method outperformed all the baseline methods in the recommendation prediction task while bridging the gap between recommendation prediction and human-interpretable explanations. Evaluations over different methods showed that item descriptions are crucial in improving performance for sparse datasets while user profiles are crucial for long user-item interaction sequences. Our experiments also illustrated how the proposed method helped in improving both recommendation prediction and reasoning generation in an SLM. Finally, we demonstrated a strong positive correlation between recommendation prediction and the quality of generated reasoning, underscoring the importance of reasoning and explanations in providing accurate recommendations.



\section*{Limitations and Future Work}
\label{section:limitation}

This paper proposes to generate synthetic explanations as a guide for LLM recommendation fine-tuning. However, how plausible the reasoning might be, the explanations generated are an approximation of the actual reasoning justifying the user's action. Hence there is a need for follow-on work to more rigorously explore the failure cases of synthetic explanations and their impact. There is also a need for further study of the interpretability of the LLM-generated explanations; do they reflect actual user motivations?

We have focused on one aspect of recommendation -- prediction of likes or dislikes while producing plausible and human-interpretable explanations. It is an open question how the method presented here could impact other recommendation tasks like candidate retrieval and re-ranking.





\bibliography{custom}

\clearpage
\appendix

\section{Experimental Setup}
\label{section:appendix_experimental_setup}
\subsection{Dataset}
\label{section:appendix_dataset}
In this study, we evaluate our methods on three recommendation datasets. The datasets vary significantly in domains,
platforms, implicitness and sparsity as shown in Table~\ref{tab:datasets_users_items}:
\paragraph{\textbf{MovieLens}}A widely used benchmark dataset. We used the version with 1 million user-item interactions, called \textit{ML1M}. This is a highly dense dataset. The item metadata, $\mathcal{M}_i$, is, however, limited to Title, Genre, and Year (of release) only and doesn't include any user review, making it highly implicit.
\paragraph{\textbf{Amazon}}A series of datasets comprising large corpora of product reviews crawled from Amazon.com. Top-level product categories on Amazon are treated as separate datasets. We consider two categories—‘Beauty’ and ‘Fashion.’ This dataset is notable for its high sparsity and variability. $\mathcal{M}_i$ consists of title, price, brand, description (only 8\% of items), and user reviews.

\begin{table}[h]
  \centering
  \resizebox{0.8\columnwidth}{!}{%
  \begin{tabular}{cccc}
    \hline
    \hline
    \textbf{Dataset} & \textbf{\#Users} & \textbf{\#Items} & \textbf{Sparsity} \\ \hline
    ML1M & 6040 & 4000 & Very Low \\
    Fashion & 1273 & 6089 & High \\
    Beauty & 624 & 1220 & Very High \\ \hline
    \end{tabular}%
  }
\caption{Datasets, their respective number of unique users and items, and sparsity.}
  \label{tab:datasets_users_items}
\end{table}

\begin{figure}[h]
  \includegraphics[width=\columnwidth]{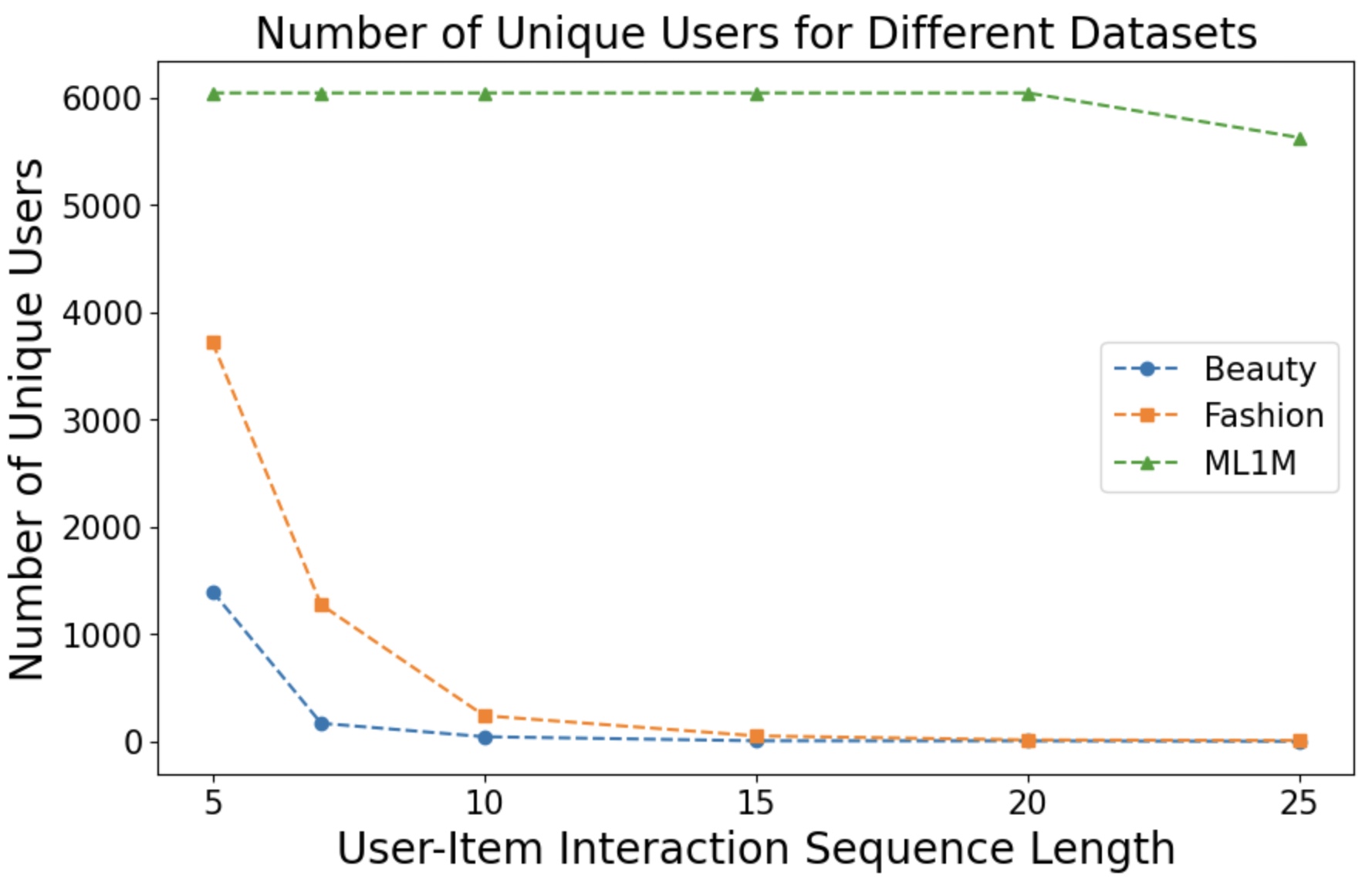}
  \caption{The multi-line graph represents the unique number of users with at least $k$ user-item interactions in each dataset. ML1M drops beyond $k=20$ suggesting very dense dataset. However, Beauty and Fashion drops drastically from $k=5$ to $k=7$, highlighting the sparsity in the dataset.}
  \label{fig:k-core}
\end{figure}


As described in Section 3, our method requires the most recent $k$ items interacted with by a user $u$ to predict recommendations. For these datasets, we define the $k$-core as the set of users with at least $k$ interactions. Following the approach of baseline methods such as SASRec, Caser, and TALLRec, we set $k = 5$ for the Beauty and Fashion datasets, and $k = 20$ for ML1M. 
Figure~\ref{fig:k-core} illustrates the sparsity of the datasets in terms of the number of unique users relative to the $k$ values in corresponding k-core datasets.

For creating the Test, Valid, and Train dataset, we follow the same approach as SASRec, and split the sequence $\mathcal{S}^u$ for each user u into three parts:
\begin{itemize}
    \item For Test data, we consider the most recent interaction $s_{|\mathcal{S}|}^u$ as $i_t$.
    \item For Validation data, we consider the second most recent interaction $s_{|\mathcal{S}|-1}^u$ as $i_t$.
    \item For Training data, we consider the third most recent interaction $s_{|\mathcal{S}|-2}^u$ as $i_t$.
\end{itemize}
and their corresponding historical sequences $(s_i^u)_{i=t-k-1}^{t-1}$ are used as past user-item interaction sequence for recommendation prediction and reasoning generation.

\subsection{Baselines}
\label{section:appendix_baselines}
To demonstrate the effectiveness of our methods, we include three different baselines:
\paragraph{\textbf{SASRec}} is an attention-based sequential recommender that employs a multi-headed self-attention architecture, similar to the Transformer decoder, to capture user preferences. In the final layer of the network, we use the Sigmoid activation function instead of Softmax, as we are performing binary classification.
\paragraph{\textbf{Caser}} is a CNN-based sequential recommender that embeds a sequence of recent items as an image and learns sequential patterns using horizontal and vertical convolution filters. Similar to SASRec, in the final layer, we use the Sigmoid activation function instead of Softmax, due to the binary classification task.
\paragraph{\textbf{TALLRec}} proposes an efficient tuning framework to align LLMs with recommendation tasks. Initially, they instruction-tune LLaMA-7B using the Alpaca dataset, followed by additional instruction finetuning with a recommendation dataset. In our study, we use \textit{Llama-2-7b-chat-hf}\footnote{Llama 2 Community License Agreement}, denoted as $\mathcal{L}_L$, which is already instruction-tuned. Thus, we focus on instruction fine-tuning with the recommendation dataset directly.
\paragraph{\textbf{Rec-SAVER}} addresses the simultaneous tasks of rating prediction and reasoning generation. It finetunes a Flan-T5 XL model to generate prediction and reasoning. It employs a highly computationally intensive training method, by generating multiple reasoning and self-validating the reasoning outputs to select the best reasoning for training. We were unable to reproduce their results on our dataset due to limited computational resources. Instead, we compare results on the Beauty dataset using the results reported in their paper.

\subsection{Evaluation Metrics}
\label{section:appendix_evaluation_metrics}
Our study focuses on both recommendation prediction and the quality of reasoning generation. For recommendation prediction, we adopt a binary classification approach, where the LLM, $\mathcal{L}$, predicts the binary rating $\hat{r}_{u,i_t} \in \{1,0\}$, with 1 corresponding to a ``like'' and 0 to a ``dislike''. The LLM outputs a sequence in the form of \textit{Prediction: Yes/No}, where \textit{'Yes'} indicates $\hat{r}_{u,i_t} = 1$ and \textit{'No'} indicates $\hat{r}_{u,i_t} = 0$. To evaluate this binary classification, we employ \textbf{Binary AUC}.

For the reasoning generation task, we use the model $\mathcal{L}_M$ to generate ground truth reasoning $R{u, i_t}$, as described in Section 3.5.1. To compare the performance of our methods in both Zero-shot and Finetuned ReasoningRec settings, we evaluate the generated reasoning $\hat{R}{u, i_t}$ against the ground truth reasoning $R{u, i_t}$. We use \textbf{BERTScore}, which calculates token-level maximum semantic similarity, making it the most comprehensive metric for this evaluation. Specifically, we use the \textit{‘deberta-xlarge-mnli’} model as the BERTScore backbone.

\subsection{Implementation details.}
\label{section:appendix_implementation_details}
We use $r_\mathcal{T} = 3$ for all of our datasets such that we convert the original user rating $r'_{u, i}$ to binary rating $r_{u,i}$ for recommendation tasks for this study.

We use $\mathcal{L}_M$ for multiple generation tasks. The hyper-parameters used for $\mathcal{L}_M$ is listed in Table~\ref{tab:parameters_zeroshots}.
\begin{table}[h]
\centering
\resizebox{\columnwidth}{!}{%
\begin{tabular}{cccc}
\hline
\hline
\textbf{Generation Task} & \textbf{temperature} & \textbf{top\_p} & \textbf{max\_new\_tokens} \\ \hline
User Profile & 0.01 & 0.9 & 256 \\
Item Description & 0.01 & 0.9 & 64 \\
Reasoning Ground Truth & 0.01 & 0.75 & 256 \\
Zero-shot ReasoningRec & 0.01 & 0.9 & 300 \\ \hline
\end{tabular}%
}
\caption{\label{tab:parameters_zeroshots}
    $\mathcal{L}_M$ hyper-parameters used for different generation tasks in zero-shot settings.
  }
\end{table}

The \textit{temperature} hyper-parameter controls the degree of randomness in the model's output generation. We used a very low temperature throughout our experiments to ensure consistent results.

The \textit{top-p} hyper-parameter regulates the range of tokens the model considers when generating the next token. For the reasoning generation task, we selected a lower value to minimize the chances of off-topic or illogical outputs, which is crucial for instruction finetuning our SLM. This also helps to enhance the coherence and consistency of the generated reasoning.

The \textit{max\_new\_tokens} hyper-parameter limits the number of tokens the model can generate. As discussed in \textit{Section 3}, the maximum number of tokens generated is in accordance with the instructions provided to the LLM.

\subsubsection{\textbf{Item Description Generation}}
\label{section:appendix_item_desc}
Section~\ref{section:item_description_generation} introduced that we generate a concise $n$ - word item description, $D_i$ for every item $i$. For our experiments we have used $n = 25$. Given, the varied level of implicitness, we describe the process of $D_i$ generation. The \textit{ML1M} dataset, being highly implicit, provides very limited item metadata which contains only the year and genre of each movie.  To enrich this information, we utilize the \textit{Cinemagoer} Python library to retrieve the plot for each movie. We then prompt, \textit{Mixtral-8X7b-Instruct-v0.1}\footnote{Apache license 2.0}, $\mathcal{L}_{M}$ as shown in Appendix~\ref{section:appendix_prompts} Table~\ref{tab:item_description_prompt} to generate a 25-word summary for each movie.

For the \textit{Beauty} and \textit{Fashion} datasets, we observed that the item titles contain some basic information about the products such as title, brand, price, description and reviews. However, only 8\% of the items in these datasets include a description in the metadata, $\mathcal{M}_i$. To address this, for each item $i$, we leverage the ratings and reviews provided by the users, $<r'_{u,i}, K_{u,i}> \forall u\in\mathcal{U}$ and $s_i^u\in\mathcal{S}^u$, to generate a more comprehensive item description. With the aim of capturing both strong (positive) and weak (negative) features of each item $i$, we sample up to $p = 10$ reviews from the available pool of reviews. These sampled reviews are selected in a stratified manner to have the original distribution across all rating levels for the item, ensuring original rating representation. The pseudo code for selecting reviews is demonstrated in Algorithm~\ref{alg:select_reviews}. We then prompt $\mathcal{L}_{M}$, with the selected reviews following the format shown in Table~\ref{tab:item_description_prompt} and Equation~\ref{eq:item_description} to generate a concise 25-word description for each item.

\begin{algorithm}
\caption{Select Reviews for an Item}
\label{alg:select_reviews}
\begin{algorithmic}[]
\REQUIRE item\_reviews (reviews grouped by rating. $r_{u,i}\in[1,5]$ for item $i$ such that $u\in\mathcal{U}$ and $s_i^u\in\mathcal{S}^u$)
\ENSURE reviews\_for\_description (list of up to 10 sampled reviews)

\STATE total\_reviews $\leftarrow$ total\_count(item\_reviews)
\STATE rating\_wise\_reviews $\leftarrow$ count(item\_reviews)
\IF {total\_reviews $\leq$ 10}
    \STATE Select all reviews for description.
\ELSE
    \STATE Calculate rating\_distribution
    \STATE Determine rating\_count (number of reviews to sample per rating) based on distribution.
    \STATE Adjust rating\_count to ensure 10 reviews are selected.
    \STATE Randomly sample reviews for each rating based on rating\_count.
\ENDIF
\STATE Trim each review to the first 50 words.
\RETURN reviews\_for\_description
\end{algorithmic}
\end{algorithm}

\subsubsection{\textbf{Instruction Finetuning}}
\label{section:appendix_instruction_finetuning}
It is the most crucial task for the purpose of getting the best finetuned model. Given the LLMs are pre-trained on huge amount of corpus, even few training examples can improve their performance considerably. We adopted a textit{$\mathcal{K}$-shot training} paradigm, similar to TALLRec, where only a limited number of samples, $\mathcal{K}\in\{64, 128, 256\}$, are randomly selected from the training dataset for instruction finetuning. We used \textit{Llama-2-7b-chat-hf} ($\mathcal{L}_L$) as the base model for finetuning, both for TALLRec and our ReasoningRec.

For TALLRec, we used the parameters used in the original implementation to generate the results - maximum numbers of epochs = 100, learning rate = $1e-4$, LoRA r = 8, alpha = 16, dropout = 0.05, and target modules = [``q\_proj'', ``v\_proj''].

For finetuning ReasoningRec, we tuned hyperparameters from the following range - learning rate: [$1e-3, 3e-4, 1e-4, 4e-5, 5e-6$], LoRA r: [4, 8] and alpha: [8, 16]. The hyperparameters used for the best model are - max epochs = 100, LoRA r = 8, alpha = 16, dropout = 0.05, target modules = [``q\_proj'', ``v\_proj''], and learning rate = $1e-4$ for Beauty and Fashion, and $3e-5$ for ML1M. We also used the \textit{SFTTrainer} with \textit{DataCollatorForCompletionOnlyLM} and \textit{Early Stopping} scheduler with patience = 5 to train the model with the cross-entropy loss from the generation output. We use the max seq length = $2048$.

\subsubsection{\textbf{Model Quantization}}
\label{section:appendix_model_quantization}
$\mathcal{L}_M$ is used with 4-bit quantization with quant\_type = \textit{nf4} and compute\_dtype = \textit{bfloat16}. $\mathcal{L}_L$ is used with 8-bit quantization with dtype = \textit{float16}. 

\section{Additional Results}
\label{section:appendix_additional_results}

\subsection{Template Ablation}
\label{section:appendix_template_ablation}

Table~\ref{tab:template_ablation} demonstrates the consistency of our prompt for zero-shot ReasoningRec. We generated multiple prompts by tweaking instructions and validated that the models perform almost similar given they are provided with rich semantic and contextual information such as Item Description and User Profile. We used the \textit{Mixtral-8x7b-Instruct-v0.1} for all these experiments.

\begin{table}[h]
\centering
\resizebox{\columnwidth}{!}{%
\begin{tabular}{cccc}
\hline
\hline
\multirow{2}{*}{\textbf{Model}} & \textbf{ML1M} & \textbf{Fashion} & \textbf{Beauty} \\
 & \textbf{AUC} & \textbf{AUC} & \textbf{AUC} \\ \hline
\begin{tabular}[c]{@{}c@{}}Zeroshot ReasoningRec\\ (Template 1)\end{tabular} & 0.614 & 0.731 & 0.693 \\
\begin{tabular}[c]{@{}c@{}}Zeroshot ReasoningRec\\ (Template 2)\end{tabular} & 0.608 & 0.731 & 0.693 \\
\begin{tabular}[c]{@{}c@{}}Zeroshot ReasoningRec\\ (Template 3)\end{tabular} & 0.605 & 0.717 & 0.687 \\ \hline
\end{tabular}%
}
\caption{\label{tab:template_ablation}
Ablation Study with different templates
}
\end{table}

\section{Prompts and Outputs}
\label{section:appendix_prompts}
\begin{table*}
  \centering
  \resizebox{\linewidth}{!}{
  \begin{tabular}{p{2cm}p{15cm}}
    \hline
    \hline
    \textbf{Dataset} & \textbf{Prompts for Item Description Generation} \\
    \hline
    ML1M       &   You are an expert movie critic. You are provided with the Movie - \{Title, Year, Plot\}. Summarize the movie in at most 25 words. \\
    Fashion/Beauty    &  As an expert \{fashion/beauty\} product recommender and advertiser, extract the strong (positive) and weak (negative) features or characteristics of the product from the given reviews. You are given the list of reviews about the product - \{List of Reviews\}. Give a 25 word concise product description mentioning strong and weak features of the product. \\
    \hline
  \end{tabular}}
\caption{\label{tab:item_description_prompt}
   Instruction Prompts given to LLM ($\mathcal{L}_M$) for Item Description Generation. The LLM is assigned a dataset specific role to generate a concise 25-word item description $D_i$.
  }
\end{table*}

\begin{table*}
  \centering
  \resizebox{\linewidth}{!}{
  \begin{tabular}{p{2cm}p{15cm}}
    \hline
    \hline
    \textbf{Dataset} & \textbf{CoT Prompts for User Profile Generation} \\
    \hline
    ML1M       &   You are an expert movie critic. List of user's liked and disliked movies and their descriptions are given - $<r_{u,i}, D_i>_{i=1}^m$
    Generate a user profile in at most 100 words. Do not include information not present in the movie descriptions.\\
    Fashion/Beauty    &  You are an expert \{fashion/beauty\} product reviewer and recommender. You are provided with a user's list of recent products and their descriptions that the user purchases and whether the user liked it or disliked it. Please go through the list in order -
    $<r_{u,i}, D_i>_{i=1}^m$.     Analyze the provided list of products purchased by the user in order and summarize the user behavior by identifying the characteristics liked and disliked about the products in at most 100 words. Do not include information not present in the item descriptions. \\
    \hline
  \end{tabular}}
\caption{\label{tab:user_profile_prompt}
   Chain-of-Though (CoT) Prompts given to LLM ($\mathcal{L}_M$) for generating User Profile. The LLM is assigned a dataset specific role to generate a 100-word user profile $P_u$. The CoT prompt consists of user preference $r_{u,i}$ and item description $D_i$ for the first $m$ items from user's historical interactions.
  }
\end{table*}

\begin{table*}
  \centering
  \resizebox{\linewidth}{!}{
  \begin{tabular}{p{2cm}p{15cm}}
    \hline
    \hline
    \textbf{Dataset} & \textbf{CoT Prompt for Reasoning Generation} \\
    \hline
    ML1M       &   You are an expert movie critic. You are provided with a user profile and a list of recent movies that the user has watched and whether the user likes it or not. User Profile - \{$P_u$\}. User watched the following movies recently in the given order. List of recent movies and their description - \{Liked/Disliked\} \{$(D_i)_{i = t-k-1}^{t-1}$\}. Now, We know that the user will \{$r_{u,i_t}$\} the next movie - \{$D_{i_t}$\}. As an expert movie critic, do the following steps - 1. Analyze the user's movie watching history given in order and summarize the user's behavior from the given list of movies and identify the movie characteristics that he likes and dislikes in atmost 100 words. 2. Explain with reasoning why the user will \{$r_{u,i_t}$\} the target movie \{$i_t$\} in atmost 100 words.
        Do not use any information not mentioned above.\\
    Fashion/Beauty    &  You are an expert \{fashion/beauty\} product recommender. You are provided with a user's profile and list of recent products and their descriptions that the user purchased and whether the user liked it or disliked it. User Profile - \{$P_u$\}. User purchased the following items in the given order. List of recent items and their description - \{Liked/Disliked\} \{$(D_i)_{i = t-k-1}^{t-1}$\}. Now, we know that the user will \{$r_{u,i_t}$\} the next item - \{$D_{i_t}$\}. As an expert \{fashion, beauty\} product recommender, do the following steps -
    1. Analyze the user profile, list of products purchased by the user in order and summarize the user behavior by identifying the characteristics he liked and disliked about the products in at most 100 words.
    2. Explain with reasoning why the user will \{$r_{u,i_t}$\} the next item - \{$i_t$\}, in atmost 100 words.\\
    \hline
  \end{tabular}}
\caption{\label{tab:response_generation_prompts}
   Chain-of-Though (CoT) Prompts given to LLM ($\mathcal{L}_M$) for generating ground truth Reasoning $R_{u,i_t}$. The prompts consist of CoT instructions along with User Profile $P_u$, Item Description $D_i$ of the $k$ most recent user-item interactions and user preference $r_{u,i}$ for those corresponding items, and the user ground truth preference for the target item $r_{u,i_t}$  
  }
\end{table*}

\begin{table*}
  \centering
  \resizebox{\linewidth}{!}{
  \begin{tabular}{p{2cm}p{15cm}}
    \hline
    \hline
    \textbf{Dataset} & \textbf{CoT Prompts for ReasoningRec} \\
    \hline
    ML1M       &  You are an expert movie critic. You are provided with the user profile and list of recent movies that the user has watched and whether the user likes it or not.
    User Profile - \{$P_u$\}. User watched the following movies recently in the given order. List of recent movies and their description - \{Liked/Disliked\} $<D_i>_{i = t-k-1}^{t-1}$ Analyze all the information given in order. Do not use any information not mentioned above. Predict whether the user will like the target movie - \{$D_{i_t}$\} or not. Answer with a Yes or No in the following format - Prediction: Yes or No by analyzing the user's behavior from the given list of movies and identify the movie characteristics that the user likes and dislikes in at most 100 words. Explain with reasoning whether the user will like or dislike the next target movie - \{$i_t$\} in at most 100 words. \\
    Fashion/Beauty    &  You are an expert \{fashion/beauty\} product recommender. You are provided with a user's profile and list of recent products and their descriptions that the user purchases and whether the user liked it or disliked it. User Profile - \{$P_u$\}. User purchased the following items in the given order. List of recent items and their description - \{Liked/Disliked\} $<D_i>_{i = t-k-1}^{t-1}$. As an expert \{fashion/beauty\} product recommender, do the following steps - Predict whether the user will like the target item - \{$D_{i_t}$\} or not in the following format - Prediction: Yes or No by analyzing the provided list of products purchased by the user in order and summarize the user behavior by identifying the characteristics the user liked and disliked about the products in at most 100 words.
    Explain with reasoning whether the user will like or dislike the next target item - \{$i_t$\} in at most 100 words.\\
    \hline
  \end{tabular}}
\caption{\label{tab:reasoning_rec_prompt}
   Chain-of-Though (CoT) Prompts given to LLM ($\mathcal{L}_M$) for generating recommendation prediction $\hat{r}_{u,i_t}$ and reasoning $\hat{R}_{u,i_t}$. The prompts consists of CoT instructions along with User Profile $P_u$, Item Description $D_i$ of the $k$ most recent user-item interactions and user preference $r_{u,i}$ for those corresponding items, and the item description of target item $D_{i_t}$
  }
\end{table*}
\begin{table*}
  \centering
  \resizebox{\linewidth}{!}{
  \begin{tabular}{p{2cm}p{15cm}}
    \hline
    \hline
    \textbf{Method} & \textbf{Generation Output} \\
    \hline
    Reasoning Ground Truth  (Mixtral)     &  \textcolor{custom_cyan}{Prediction: Yes} 1. The user profile shows that the user \textcolor{green}{likes blouses that are cute, classy, and stylish. They prefer blouses with small neck openings and tight sleeves}. The user \textcolor{red}{dislikes blouses that are sheer, not stretchy, and run small}. \textcolor{green}{They also prefer blouses that are not too revealing and are comfortable to wear in a desk job}.  2. \textcolor{custom_purple}{The user will like the next item, Pandora Women's Charm 791726PCZ, Silver, because it is a beautiful and original charm that is perfect for Pandora charm bracelets. Although the product may not be love at first sight and may not meet all expectations, the user has shown a preference for fashion items that are stylish and attractive. The charm is a fashion accessory that can add a touch of elegance to any outfit, making it a great addition to the user's collection. Additionally, the user has shown a preference for fashion items that are not too revealing, and the charm is a subtle and elegant accessory that can be worn in any setting.} \\
    ReasoningRec (Finetuned Llama2) &  \textcolor{custom_cyan}{Prediction: Yes} 1. The user \textcolor{green}{seems to prefer blouses with lace details and a double-layered design. They also seem to prefer loose-fitting blouses}, as indicated by the user's purchase of the BLDO Women's Short Sleeve Lace Patchwork Short Loose Chiffon Blouse and BLDO Women's Summer Slim Lace Sleeve Chiffon Blouse Crop Top. The user also seems to \textcolor{green}{prefer products that are stylish and attractive}, as indicated by their purchase of the Allegra K Women Peter Pan Collar Sheer Crochet Sleeves Chiffon Blouse M Beige and Zeagoo Women's Chiffon Long Sleeve Blouse Tops Shirts Lace Embellished. 2. \textcolor{custom_purple}{The user will like the next item, Pandora Women's Charm 791726PCZ, Silver, because it is a stylish and attractive accessory that can be added to a Pandora charm bracelet. The product description states that it may not meet all expectations, but the user has shown a preference for stylish and attractive items.}"\\
    Zero-shot ReasoningRec (Mixtral) & \textcolor{custom_cyan}{Prediction: Yes}, the user will like the target item.The user seems to \textcolor{green}{like blouses that are stylish, attractive, and have lace detailing. They also prefer items that are loose or non-sleeveless}. The user has had issues with sizing in the past, so they \textcolor{green}{may prefer items that run true to size or have a more generous fit}.\textcolor{custom_purple}{The Pandora charm meets the user's preference for stylish and attractive items. While it is not a blouse, it is a fashion accessory that can add a touch of elegance to any outfit. The charm is also silver, which is a neutral color that can match with various outfits. Additionally, the charm is not too flashy, which aligns with the user's preference for subtle yet stylish items.} However, the user may be concerned about the size of the charm. The description does not provide information on the size of the charm, which may cause uncertainty for the user. To address this concern, the seller can provide more information on the size of the charm or offer a size chart for customers to refer to.Overall, the user's past purchases suggest that they prefer stylish and attractive items with lace detailing. The Pandora charm meets these criteria and is likely to be appreciated by the user."\\
    Zero-shot ReasoningRec (Llama2) & \textcolor{custom_cyan}{Prediction: Yes}, the user will like the target item. Liked Allegra K Women Peter Pan Collar Sheer Crochet Sleeves Chiffon Blouse M Beige Brand: Allegra K Description: The product is a cute, classy blouse with a small neck opening and tight sleeves. It is available in cream and black, but the colors may vary from the picture. The quality is average and it may cause discomfort in a desk job.Liked Zeagoo Women's Chiffon Long Sleeve Blouse Tops Shirts Lace Embellished, apricot, Asian XL(US L) Brand: Zeagoo Description: The product is a beautiful blouse with pearls on the top and detailed design. However, it runs small, is sheer, and not stretchy, causing fitting issues.Liked BLDO Women's Short Sleeve Lace Patchwork Short Loose Chiffon Blouse White (XXL) Brand: BLDO Description: The product is a pretty, double-layered, non-sleeveless blouse with lace detailing. However, it runs significantly smaller than the listed size, requiring customers to order 2 sizes up."\\
    \hline
  \end{tabular}}
\caption{\label{tab:fashion_example_output}
   Generated outputs for User - 46 Target Item - Pandora Women's Charm 791726PCZ in Amazon Fashion: ReasoningRec and Zero-shot ReasoningRec (Mixtral) provide similar reasoning as the Reasoning Ground Truth. However, Zero-shot ReasoningRec (Llama2) just repeats the input prompt without providing any reasoning.
  }
\end{table*}
\end{document}